\documentclass[letterpaper]{mn2e}
\usepackage{rotating}
\usepackage{colortbl}
\usepackage[fleqn]{amsmath}
\usepackage{amsfonts}

\newcommand{\apj}{ApJ}
\newcommand{\apjl}{ApJ}
\newcommand{\apjs}{ApJS}
\newcommand{\aap}{A\&A}
\newcommand{\aj}{AJ}
\newcommand{\mnras}{MNRAS}

\newcommand{\nat}{Nat}

\onecolumn

\title[Hydrostatic equilibrium profiles in elliptical galaxies]{Hydrostatic equilibrium profiles for gas in elliptical galaxies}

\author[P.R. Capelo, P. Natarajan and P.S. Coppi]{Pedro R. Capelo$^{1}$\thanks{e-mail: pedro.capelo@yale.edu}, Priyamvada Natarajan$^{1,2}$ and Paolo S. Coppi$^{1,2}$\\
$^1$Department of Astronomy, Yale University, P.O. Box 208101, New Haven, CT 06520-8101, USA\\
$^2$Department of Physics, Yale University, P.O. Box 208120, New Haven, CT 06520-8120, USA}

\begin{document}
\maketitle

\begin{abstract}
We present an analytic formulation for the equilibrium gas density profile of early-type galaxies that explicitly includes the contribution of stars in the gravitational potential. We build a realistic model for an isolated elliptical galaxy and explore the equilibrium gas configurations as a function of multiple parameters. For an assumed central gas temperature $k_BT_0=0.6$ keV, we find that neglecting the gravitational effects of stars, which can contribute substantially in the innermost regions, leads to an underestimate of the enclosed baryonic gas mass by up to $\sim$65\% at the effective radius, and by up to $\sim$15\% at the NFW scale radius, depending on the stellar baryon fraction. This formula is therefore important for estimating the baryon fraction in an unbiased fashion. These new hydrostatic equilibrium solutions, derived for the isothermal and polytropic cases, can also be used to generate more realistic initial conditions for simulations of elliptical galaxies. Moreover, the new formulation is relevant when interpreting X-ray data. We compare our composite isothermal model to the standard $\beta$-model used to fit X-ray observations of early-type galaxies, to determine the value of the NFW scale radius $r_s$. Assuming a 10\% stellar baryon fraction, we find that the exclusion of stars from the gravitational potential leads to (i) an underestimate of $r_s$ by $\sim$80\%, and to (ii) an overestimate of the enclosed dark matter at $r_s$ by a factor of $\sim$2, compared to the equivalent $\beta$-model fit results when stars are not taken into account. For higher stellar mass fractions, a $\beta$-model is unable to accurately reproduce our solution, indicating that when the observed surface brightness profile of an isolated elliptical galaxy is found to be well fitted by a $\beta$-model, the stellar mass fraction cannot be much greater than $\sim$10\%.

\end{abstract}


\section{Introduction}

Several studies, both observational and computational, of groups and clusters of galaxies, and of individual early-type galaxies, assume that the hot gas present in these systems is in hydrostatic equilibrium (HE) in the overall gravitational potential. Even though the assumption of HE has been shown to not always be robust (e.g. Bertin et al. 1993; Ciotti \& Pellegrini 2004), many authors assume an isothermal $\beta$-model (Cavaliere \& Fusco-Femiano 1976, 1978) to describe the gas profile, but several solve the HE equation, assuming that the gravitational potential is due to dark matter (DM) only, usually described by an NFW (Navarro, Frenk \& White 1996) profile (e.g. Makino, Sasaki \& Suto 1998; Suto, Sasaki \& Makino 1998), or by more general profiles (e.g. Ciotti \& Pellegrini 2008). While this approximation seems to hold for groups and clusters of galaxies, it is well known that stars contribute a significant fraction of the total and baryonic matter of elliptical galaxies, and even become the dominant part at the very centre of these systems (e.g. Ferreras, Saha \& Williams 2005). The purpose of this paper is to extend previous calculations that only considered the potential due to DM, and explicitly include the contribution of the stellar component as well.

Providing an analytic fitting formula for calculating the response of the gas to the presence of the gravitational potential induced by both DM and stars is useful while interpreting X-ray data of early-type galaxies and for setting up initial conditions for simulations of elliptical galaxies. The derivation of an equilibrium gas profile neglecting the effects of stars leads to an incorrect estimate of the enclosed gas mass, particularly in the central regions of the galaxy, and therefore has implications for the evaluation of the baryon fraction as a function of radius, and for the calculation of the rate of accretion onto a central supermassive black hole. Moreover, the known relations (e.g. Makino et al. 1998) between observed properties of the hot gas, described via a $\beta$-model, and the parameters of the underlying DM gravitational potential, are significantly affected by the inclusion of the stellar component. In addition, an analytic formula that includes the effects of stars can be important to provide a more realistic formulation of the initial equilibrium conditions for simulations of galaxy formation and evolution (e.g. Di Matteo, Springel \& Hernquist 2005; Springel, Di Matteo \& Hernquist 2005; Hopkins et al. 2006).

It has long been observed, both at low redshift (e.g. Graham et al. 1996; Kormendy et al. 2009) and high redshift (e.g. Mancini et al. 2010; van Dokkum et al. 2010), that the surface brightness of many elliptical galaxies can be well described by the empirically derived de Vaucouleurs (1948) profile or, more generally, by the S{\'e}rsic (1963, 1968) profile. Unfortunately, the de-projected stellar mass density profile cannot be calculated analytically (but see Young 1976 for a numerical study). There is, however, a family of stellar mass density profiles (Dehnen 1993; see also Tremaine et al. 1994) that also describes observed properties well. Not only is the Dehnen family analytically simple, but also, for a defined range of parameters, it quite successfully resembles the de Vaucouleurs profile in both surface density and distribution function. Furthermore, it includes, as special cases, the widely used Jaffe (1983) and Hernquist (1990) profiles. It is this family of density profiles that we investigate further in this paper, and use to model the stellar component. Throughout the paper, we use the following values for cosmological parameters when needed: $\Omega_0=0.27$, $\Omega_b=0.045$, $\Omega_{\Lambda}=0.73$, $H_0=100 \,h$ km s$^{-1}$ Mpc$^{-1}=70 \,h_{70}$ km s$^{-1}$ Mpc$^{-1}$, $h=0.7$ (i.e. $h_{70}=1.0$).

The outline of the paper is as follows. In Section \ref{sec:Formulation}, we formulate the problem and derive the new formalism for HE gas profiles in the presence of DM and stars. In Section \ref{sec:Analysis}, we analyse the HE solutions for a realistic model of an isolated elliptical galaxy, compare the results for a wide range of parameters, and present new relations between $\beta$-model and DM parameters. We conclude in Section \ref{sec:Conclusions} with a discussion of our results and their implications for interpreting X-ray observations, and for simulations of elliptical galaxies.

\section{Formulation}\label{sec:Formulation}

We model the composite system of gas in equilibrium with stars and a DM halo. To start with, we consider the case of an ideal gas in a generic total gravitational potential $\phi$ with a finite central value $\phi_0$, and solve the HE equation, $\nabla P = - \rho \nabla \phi$, where $P$ and $\rho$ refer to the pressure and mass density of the gas. Note that the general equation simplifies in the case of spherical symmetry to the familiar form

\begin{equation}
\frac{dP}{dr} = - \rho \frac{d\phi}{dr} = - \rho \frac{G M_T(r)}{r^2},
\end{equation}

\noindent where we have made use of Poisson's equation, $\nabla^2\phi=4\pi G \rho_T$, and $M_T(r)$ and $\rho_T(r)$ are the total enclosed mass and total mass density, respectively. If the gas is barotropic, i.e. $\rho=\rho(P)$, we can further write

\begin{equation}
\int_{P_0}^P{\frac{1}{\rho(P')}dP'} = -\int_{\phi_0}^{\phi}{d\phi '} = - G \int_0^r{\frac{M_T(r')}{r'^2}\,dr'},
\end{equation}

\noindent where the free parameter $P_0=P(0)$ was chosen as the boundary condition.

For simplicity, we only consider the two common barotropic cases: the isothermal and the polytropic case. In the isothermal ideal gas case, pressure and density are related by $P(r) = [k_B T_0/(\mu m_p)] \rho(r) = (P_0/\rho_0) \rho(r)$, where $k_B$ is the Boltzmann constant, $\mu$ is the gas mean molecular weight, $m_p$ is the proton mass, and $T_0$ and $\rho_0$ are the central gas temperature and density, respectively. The solution to the HE equation in this case can be reduced to the familiar integral:

\begin{equation}
\rho(r) = \rho_0\exp\left[\Delta \left(\frac{\phi}{\phi_0}-1\right)\right] = \rho_0\exp\left[-\frac{G\rho_0}{P_0} \int_0^r{\frac{M_T(r')}{r'^2}\,dr'}\right],
\label{eq:iso_sol}
\end{equation}

\noindent where $\Delta = -\phi_0 \rho_0/P_0$.

For the polytropic ideal gas case, pressure and density are related by $P(r) = K_0 \rho(r)^{\Gamma}$, where $K_0=P_0/\rho_0^{\Gamma}=k_B T_0/(\mu m_p \rho_0^{\Gamma -1})$ and $\Gamma \neq 1$ is the polytropic index.  The solution to the HE equation in this case is:

\begin{equation}
\rho(r) = \rho_0 \left[1+\frac{\Gamma-1}{\Gamma}\Delta\left(\frac{\phi}{\phi_0}-1\right)\right]^{\frac{1}{\Gamma-1}} = \rho_0 \left[1-\frac{\Gamma-1}{\Gamma}\frac{G \rho_0}{P_0}\int_0^r{\frac{M_T(r')}{r'^2}\,dr'}\right]^{\frac{1}{\Gamma-1}}.
\label{eq:poly_sol}
\end{equation}

It is worth pointing out that the polytropic solution is only physically meaningful if the term in square brackets of the r.h.s of equation \eqref{eq:poly_sol} is non-negative. Moreover, the isothermal quantities (and, in most cases, the polytropic quantities) are non-zero at infinity: with our choice of the boundary condition, regardless of the slope of the gravitational potential at large radii, only a potential which is infinite at the origin would be able to cause the gas density at infinity to be zero. For this reason, results at large radii should be treated with caution. See Bulbul et al. (2009) for an alternate approach in the polytropic case, where they impose $T(+\infty)=0$ as the boundary condition. This is equivalent to setting $(\Gamma-1)\Delta/\Gamma=1$ in equation \eqref{eq:poly_sol}.

We further assume that gas does not contribute to the total gravitational potential (i.e. we neglect gas self-gravity), and therefore solve the simplified HE equation instead, where now $\phi$ is the gravitational potential due to all mass \textit{excluding} gas, and $M_T$ is the total mass \textit{excluding} gas. This assumption is supported, especially for the inner regions, by previous studies (e.g. Suto et al. 1998), and is justified by the fact that gas is \textit{globally} negligible, i.e. the enclosed gas mass is one or more orders of magnitude smaller than the total enclosed mass for all radii up to the virial radius. Stars, on the other hand, while having a virial enclosed mass comparable to that of the gas, are not negligible in the inner parts of the galaxy: in fact, they represent the dominant matter fraction at least up to the effective radius (e.g. Ferreras et al. 2005). We initially consider a system with gas and DM only, and assume that the DM density can be described by an NFW profile,

\begin{equation}
\rho_{NFW}(r) = \frac{\delta_c \rho_c}{(r/r_s)(1+r/r_s)^2},
\label{eq:NFW_rho}
\end{equation}

\noindent where $r_s$ is a scale radius, $\delta_c$ is a characteristic (dimensionless) density, and $\rho_c$ is the critical density of the Universe. The corresponding gravitational potential and enclosed mass are

\begin{equation}
\phi_{NFW}(r) = \phi_{NFW0} \frac{\ln\left(1+r/r_s\right)}{r/r_s}, \hspace{0.5cm} M_{NFW}(r) = 4 \pi \delta_c \rho_c r_s^3 \left[\ln\left(1+\frac{r}{r_s}\right)-\frac{r}{r+r_s}\right],
\label{eq:NFW_pot_and_mass}
\end{equation}

\noindent respectively, where $\phi_{NFW0}=\phi_{NFW}(0)=- 4\pi G \delta_c \rho_c r_s^2$, and we have assumed $\phi_{NFW}(+\infty)=0$. Substituting these into equations \eqref{eq:iso_sol} and \eqref{eq:poly_sol}, we have, as in Makino et al. (1998) for the isothermal case,

\begin{equation}
\rho(r) = \rho_0 \exp\left[-\Delta_{NFW}\left(1-\frac{\ln(1+r/r_s)}{r/r_s}\right)\right],
\label{eq:Makino_sol}
\end{equation}

\noindent and, as in Suto et al. (1998) for the polytropic case,

\begin{equation}
\rho(r) = \rho_0 \left[1-\frac{\Gamma-1}{\Gamma}\Delta_{NFW}\left(1-\frac{\ln(1+r/r_s)}{r/r_s}\right)\right]^{\frac{1}{\Gamma-1}},
\label{eq:Suto_sol}
\end{equation}

\noindent where $\Delta_{NFW}=-\phi_{NFW0}\rho_0/P_0=4\pi G \delta_c \rho_c r_s^2 \mu m_p/(k_B T_0)$.

Now we move on to the case of an NFW DM halo with the stellar component modelled using a Dehnen profile, neglecting the effects of adiabatic contraction. The Dehnen (1993; see also Tremaine et al. 1994) family of stellar mass density profiles is given by

\begin{equation}
\rho_D(r) = \frac{(3 - \psi)M_*}{4 \pi} \frac{r_*}{r^{\psi}(r+r_*)^{4-\psi}},
\label{eq:Dehnen_rho}
\end{equation}

\noindent where $0 \le \psi < 3$, $M_*$ is the total stellar mass (i.e. integrating from the origin to infinity), and $r_*$ is a scale radius. The Dehnen profile is useful as its projected distribution accurately matches the de Vaucouleurs (1948) surface density profile which is an excellent fit to observations. This is especially true for values of $\psi$ ranging between 1 (which corresponds to the Hernquist profile) and 2 (the Jaffe profile), with $\psi=3/2$ being the value for which the match is closest. The Dehnen density profile corresponds to a gravitational potential

\begin{equation}
\phi_D(r) = \frac{GM_*}{r_*}
         \begin{cases}\displaystyle{\frac{1}{\psi-2} \left[1-\left(\frac{r}{r+r_*}\right)^{2-\psi}\right]} & \mbox{if $\psi \neq 2$,} \cr
                      \displaystyle{\ln\left(\frac{r}{r+r_*}\right)} & \mbox{if $\psi = 2$,} \cr
         \end{cases}
\end{equation}

\noindent where we have assumed $\phi_D(+\infty)=0$. Since equations \eqref{eq:iso_sol} and \eqref{eq:poly_sol} require a finite central gravitational potential, we only consider Dehnen profiles with $0 \le \psi < 2$, for which the gravitational potential and enclosed mass are

\begin{equation}
\phi_D(r) = \phi_{D0} \left[1-\left(\frac{r}{r+r_*}\right)^{2-\psi}\right], \hspace{0.5cm} M_D(r) = M_* \left(\frac{r}{r+r_*}\right)^{3-\psi},
\label{eq:Dehnen_pot_and_mass}
\end{equation}

\noindent respectively, where $\phi_{D0}=\phi_D(0)=-G M_*/[r_*(2-\psi)]$.

We consider a system that consists of an NFW DM halo and stars described by a Dehnen profile, for which the total gravitational potential and total enclosed mass are given by explicitly adding the contributions of the stars and DM given in equations \eqref{eq:Dehnen_pot_and_mass} and \eqref{eq:NFW_pot_and_mass}, respectively. Using equations \eqref{eq:iso_sol} and \eqref{eq:poly_sol}, we obtain

\begin{equation}
\rho(r) = \rho_0 \exp\left[-\Delta_{NFW}\left(1-\frac{\ln(1+r/r_s)}{r/r_s}\right)\right] \exp\left[-\Delta_D\left(\frac{r/r_*}{1+r/r_*}\right)^{2-\psi}\right]
\label{eq:Capelo_iso}
\end{equation}

\noindent for the isothermal case, and

\begin{equation}
\rho(r) = \rho_0 \left[1-\frac{\Gamma-1}{\Gamma}\Delta_{NFW}\left(1-\frac{\ln(1+r/r_s)}{r/r_s}\right)-\frac{\Gamma-1}{\Gamma}\Delta_D\left(\frac{r/r_*}{1+r/r_*}\right)^{2-\psi}\right]^{\frac{1}{\Gamma-1}}
\label{eq:Capelo_poly}
\end{equation}

\noindent for the polytropic case, where $\Delta_D=-\phi_{D0}\rho_0/P_0=G M_* \mu m_p/[r_* k_B T_0 (2-\psi)]$. Notice that equations (\ref{eq:Capelo_iso}-\ref{eq:Capelo_poly}) reduce to equations (\ref{eq:Makino_sol}-\ref{eq:Suto_sol}) when $M_*=0$.

\section{Analysis}\label{sec:Analysis}

\subsection{Exploring parameters and realistic matter distributions}

In this section, for a system with an NFW DM halo and stars described by a Dehnen profile, we compare equilibrium gas density profiles as a function of $f_{star}$, the stellar mass fraction at the virial radius, $\psi$, the Dehnen parameter, and $T_0$, the central gas temperature. We also consider the case when stars are included in the total mass budget, but not in the total gravitational budget.

We consider a system at redshift $z$, which consists of an isolated elliptical galaxy at the centre of an isolated, collapsed DM halo, wherein the density is given by

\begin{equation}
\rho_{NFW}(r) = \frac{M_{NFW}(r_{vir})}{4 \pi f(c)} \frac{1}{r(r+r_s)^2},
\end{equation}

\noindent where $M_{NFW}(r_{vir})$ is the DM mass enclosed within the virial radius $r_{vir}$, and $f(c_{vir}) = \ln(1+c_{vir})-c_{vir}/(1+c_{vir})$ is a function of the concentration parameter $c_{vir} = r_{vir}/r_s$, which depends on redshift and (total) virial mass $M_{vir}$ as

\begin{equation}
c_{vir} = \frac{c_0}{1+z}\left(\frac{M_{vir}}{1.3 \times 10^{13} h^{-1} M_{\odot}}\right)^{\alpha},
\end{equation}

\noindent where $c_0$ and $\alpha$ are constants that can be inferred from simulations (e.g. Bullock et al. 2001; Hennawi et al. 2007) and observations (e.g. Comerford \& Natarajan 2007; Mandelbaum, Seljak \& Hirata 2008), and $M_{vir}$ is in units of M$_{\odot}$.

The virial radius is, by definition, the radius of a spherical region within which the mean (total) mass density is $\Delta_{vir}$ times the mean critical density\footnote{We use $\rho_c(z)$; other authors, e.g. NFW, use $\rho_c(0)$.}:

\begin{equation}
r_{vir} = \left(\frac{3 M_{vir}}{4 \pi \Delta_{vir} \rho_c}\right)^{1/3},
\end{equation}

\noindent where the virial overdensity $\Delta_{vir}(z) \simeq 18 \pi^2 + 82(\Omega_m(z)-1) - 39(\Omega_m(z)-1)^2$ (Bryan \& Norman 1998)\footnote{With the above definitions, $\Delta_{vir}(0) \simeq 97$, $\Delta_{vir}(0.5) \simeq 133$, and $\Delta_{vir}(+\infty) = 18\pi^2$. Note that other authors (e.g. Bullock et al. 2001) use equivalent but different definitions, $M_{vir} = 4 \pi r_{vir}^3 \Delta_{vir} \Omega_m(z) \rho_c/3$ and $\Delta_{vir}(z) \simeq [18 \pi^2 + 82(\Omega_m(z)-1) - 39(\Omega_m(z)-1)^2]/\Omega_m(z)$, so that $\Delta_{vir}(0) \simeq 359$.}, the matter fraction $\Omega_m(z) = \Omega_0(1+z)^3/[\Omega_0(1+z)^3+\Omega_{\Lambda}]$, and the mean critical density $\rho_c(z) = 3 H_0^2 [\Omega_0(1+z)^3 + \Omega_{\Lambda}]/(8 \pi G)$.

The NFW density profile function can be also re-written as in equation \eqref{eq:NFW_rho}, with $\delta_c = f_{DM} \Delta_{vir} c_{vir}^3/(3 f(c_{vir}))$, and $f_{DM} = M_{NFW}(r_{vir})/M_{vir} = 1-f_b = 1-b_b \Omega_b/\Omega_0$, where $f_{DM}$ and $f_b$ are the DM and baryon fraction, respectively, within the virial radius, $b_b$ is the baryon fraction relative to the universal value, and we have assumed that the universal baryon fraction, $\Omega_b/\Omega_0$, is independent of redshift.

The elliptical galaxy stellar density profile is given by equation \eqref{eq:Dehnen_rho}. The total stellar mass $M_*$ is calculated by setting $r=r_{vir}$ in equation \eqref{eq:Dehnen_pot_and_mass} and by imposing\footnote{Notice that by imposing the same stellar virial mass (i.e. by fixing $f_{star}$) and by varying the Dehnen parameter, the total stellar mass $M_*$ will necessarily vary, albeit only very slightly.} $M_D(r_{vir})=f_{star}M_{vir}$. The virial stellar mass fraction can be written, following e.g. Mamon \& {\L}okas (2005b), as

\begin{equation}
f_{star} = \frac{\Upsilon_{*,B} L_{*,B}(r_{vir})}{\Upsilon_B L_{*,B}(r_{vir})} = \frac{\Upsilon_{*,B}}{b_{\Upsilon}\overline{\Upsilon}_B},
\end{equation}

\noindent where $\Upsilon_{*,B}$ is the stellar mass-to-light ratio, $\overline{\Upsilon}_B$ is the universal mass-to-light ratio, $\Upsilon_B$ is the galactic mass-to-light ratio (all in M$_{\odot}/$L$_{\odot}$), and $b_{\Upsilon}$ is the (dimensionless) mass-to-light ratio bias, all in the B-band. The scale radius $r_*$ is related to the effective radius $r_e$ through a numerical fit (Dehnen 1993),

\begin{equation}
\frac{r_*}{r_e} \simeq \frac{2^{1/(3-\psi)}-1}{0.7549-0.00439\psi+0.00322\psi^2-0.00182\psi^3},
\end{equation}

\noindent whereas the effective radius itself can be derived from empirical fits to observations\footnote{Notice that $r_e$ does not depend on the stellar fraction.} (Mamon \& {\L}okas 2005a) as $\log(h_{70}r_e) = 0.34 + 0.54 \log L_{10} + 0.25 (\log L_{10})^2$, where $L_{10} = h_{70}^2 L_B/(10^{10} L_{\odot})$, $L_B=M_{vir}/(b_{\Upsilon}\overline{\Upsilon}_B)$, and $L_B$ and $M_{vir}$ are in units of L$_{\odot}$ and M$_{\odot}$, respectively.

Finally, we can calculate the central gas density $\rho_0$ by imposing

\begin{equation}
\int_0^{r_{vir}}{\rho(r)4\pi r^2 \,dr} = f_{gas}M_{vir} = (1-f_{star}-f_{DM})M_{vir} = \left(b_b \frac{\Omega_b}{\Omega_0} - \frac{\Upsilon_{*,B}}{b_{\Upsilon}\overline{\Upsilon}_B}\right)M_{vir},
\end{equation}

\noindent where we have assumed that all matter is comprised of DM, gas, and stars only (i.e. we have neglected other forms of matter, such as dust, and the presence of a supermassive central black hole), and that $\rho(r)$ is given by equation \eqref{eq:Capelo_iso} or \eqref{eq:Capelo_poly} when stars are included in the gravitational potential, and by equation \eqref{eq:Makino_sol} or \eqref{eq:Suto_sol} when stars are excluded. Note, however, that the gas virial mass is the same with and without the stellar component in the gravitational potential, since stars are included in the total mass budget in both cases.

\renewcommand{\arraystretch}{1.3}
\begin{table}
\centering
\scalebox{0.95}{
\begin{tabular}{|p{6.5cm}|p{3cm}|p{3cm}}
\hline
\multicolumn{1}{| >{\columncolor[rgb]{0.8,0.8,0.8}}l|}{Cosmological parameters} & \multicolumn{2}{c|}{$\Omega_0=0.27$, $\Omega_{\Lambda}=0.73$, $\Omega_b=0.045$, $h=0.7$, $b_b=1$} \\
\hline
\multicolumn{1}{| >{\columncolor[rgb]{0.8,0.8,0.8}}l|}{Stellar parameters} & \multicolumn{2}{c|}{$b_{\Upsilon}=0.25641$, $\overline{\Upsilon}_B=390$ M$_{\odot}/$L$_{\odot}$, $\Upsilon_{*,B}=6.5$ M$_{\odot}/$L$_{\odot}$, $\psi=1$} \\
\hline
\multicolumn{1}{| >{\columncolor[rgb]{0.8,0.8,0.8}}l|}{Gas parameters} & \multicolumn{2}{c|}{$k_BT_0=0.6$ keV, $\mu=0.62$, $\Gamma=5/3$} \\
\hline
\multicolumn{1}{| >{\columncolor[rgb]{0.8,0.8,0.8}}l|}{Virial mass and redshift} & \multicolumn{2}{c|}{$M_{vir}=10^{12}$ M$_{\odot}$, $z=0.5$} \\
\hline
\multicolumn{1}{| >{\columncolor[rgb]{0.8,0.8,0.8}}l|}{Concentration parameters} & \multicolumn{2}{c|}{$c_0=9$, $\alpha=-0.13$} \\
\hline
Gas central density & \multicolumn{2}{c|}{$\rho_0=5.80 \times 10^{-27}$ g cm$^{-3}$} \\
\hline
Gas central pressure & \multicolumn{2}{c|}{$P_0=5.37 \times 10^{-12}$ g cm$^{-1}$ s$^{-2}$} \\
\hline
Matter fraction, critical density, and virial overdensity at redshift $z$ & \multicolumn{2}{c|}{$\Omega_z=0.56$, $\rho_c=1.51 \times 10^{-29}$ g cm$^{-3}$, $\Delta_{vir}=133$} \\
\hline
Virial radius and NFW parameters & \multicolumn{2}{c|}{$r_{vir}=200$ kpc, $r_s=22.8$ kpc, $c_{vir}=8.77$, $\delta_c=1.81 \times 10^4$} \\
\hline
Central gravitational potentials & \multicolumn{2}{c|}{$\phi_{0-NFW}=-1.14 \times 10^{15}$ cm$^2$ s$^{-2}$, $\phi_{0-D}=-2.35 \times 10^{15}$ cm$^2$ s$^{-2}$} \\
\hline
Shape parameters & \multicolumn{2}{c|}{$\Delta_{NFW}=1.23$, $\Delta_D=2.53$} \\
\hline
Stellar scale radii & \multicolumn{2}{c|}{$r_e=2.19$ kpc, $r_*=1.21$ kpc} \\
\hline
\end{tabular}
}
\caption{Relevant parameters used in our fiducial model, for an isothermal gas density profile. Given quantities are in grey cells; calculated quantities are in white cells.}
\label{tab:Parameters}
\end{table}
\renewcommand{\arraystretch}{1.0}

We note that a very small number of parameters are needed to compute the equilibrium profiles for the gas. These are, for the DM halo system, the redshift $z$, the (total) virial mass $M_{vir}$ of the system, the fitting parameters $c_0$ and $\alpha$ of the empirical relation between concentration and total virial mass, and the baryon fraction $b_b$ relative to the universal value. For the stars, the parameters are the galactic B-band mass-to-light ratio $\Upsilon_B$ (or, alternatively, the universal B-band mass-to-light ratio $\overline{\Upsilon}_B$ and the B-band mass-to-light ratio bias $b_{\Upsilon}$), the stellar B-band mass-to-light ratio $\Upsilon_{*,B}$, and the Dehnen parameter $\psi$. Finally, for the gas, the parameters are the mean molecular weight $\mu$, the polytropic index $\Gamma$, and the central temperature $T_0$ (or, alternatively, the ratio between central density and pressure $\rho_0/P_0$).

In our fiducial model, we use the Mamon \& {\L}okas (2005b) mass-to-light ratio values and choose $b_{\Upsilon}=0.25641$, $\overline{\Upsilon}_B=390$ M$_{\odot}/$L$_{\odot}$ (so that $\Upsilon_B=100$ M$_{\odot}/$L$_{\odot}$ precisely), and $\Upsilon_{*,B}=6.5$ M$_{\odot}/$L$_{\odot}$. As for the halo parameters, we choose a system with universal baryon fraction ($b_b=1$) at redshift $z=0.5$, a (total) virial mass $M_{vir}=10^{12}$ M$_{\odot}$, and the $c_{vir}$-$M_{vir}$ relation parameters $c_0=9$ and $\alpha=-0.13$ (Bullock et al. 2001). Finally, we fix the central temperature of the gas to be $k_BT_0=0.6$ keV (Diehl \& Statler 2008), the mean molecular weight of the gas to be $\mu=0.62$, the polytropic index to be $\Gamma=5/3$, and the Dehnen parameter to be $\psi=1$ (Hernquist 1990). All these quantities are listed in Table \ref{tab:Parameters}, in addition to the parameters calculated throughout this section.

The choice of the central temperature $T_0$  is particularly important, since it greatly influences the values of $\Delta_{NFW}$ and $\Delta_D$. If we use the (cluster and group) mass-temperature (M-T) relation (e.g. Sanderson et al. 2003) or, rather, extrapolate the M-T relation down to the mass ranges typical of early-type galaxies, we obtain e.g. $k_BT_0 \simeq 0.2$ keV for a system with $M_{vir}=10^{12}$ M$_{\odot}$. However, the extrapolation of the M-T relation is not without risk, as there is some evidence for a gradual steepening of  the relation, with decreasing mass (Sanderson et al. 2003). Using individual observations of isolated elliptical galaxies, the central temperatures measured are usually slightly higher, ranging from $k_BT_0 \sim 0.5$ keV (e.g. Memola et al. 2009) to $k_BT_0 \sim 1$ keV (e.g. O'Sullivan \& Ponman 2004; O'Sullivan, Sanderson \& Ponman 2007). Finally, Diehl \& Statler (2008) give a convincing argument for a central temperature $k_BT_0 \sim 0.6$ keV, using observations of 36 normal early-type galaxies. For this reason, here we study cases with three different central temperatures: $k_BT_0=0.2,0.6,1$ keV.

\begin{figure}
\begin{center}
\includegraphics[width=11.5cm,angle=90]{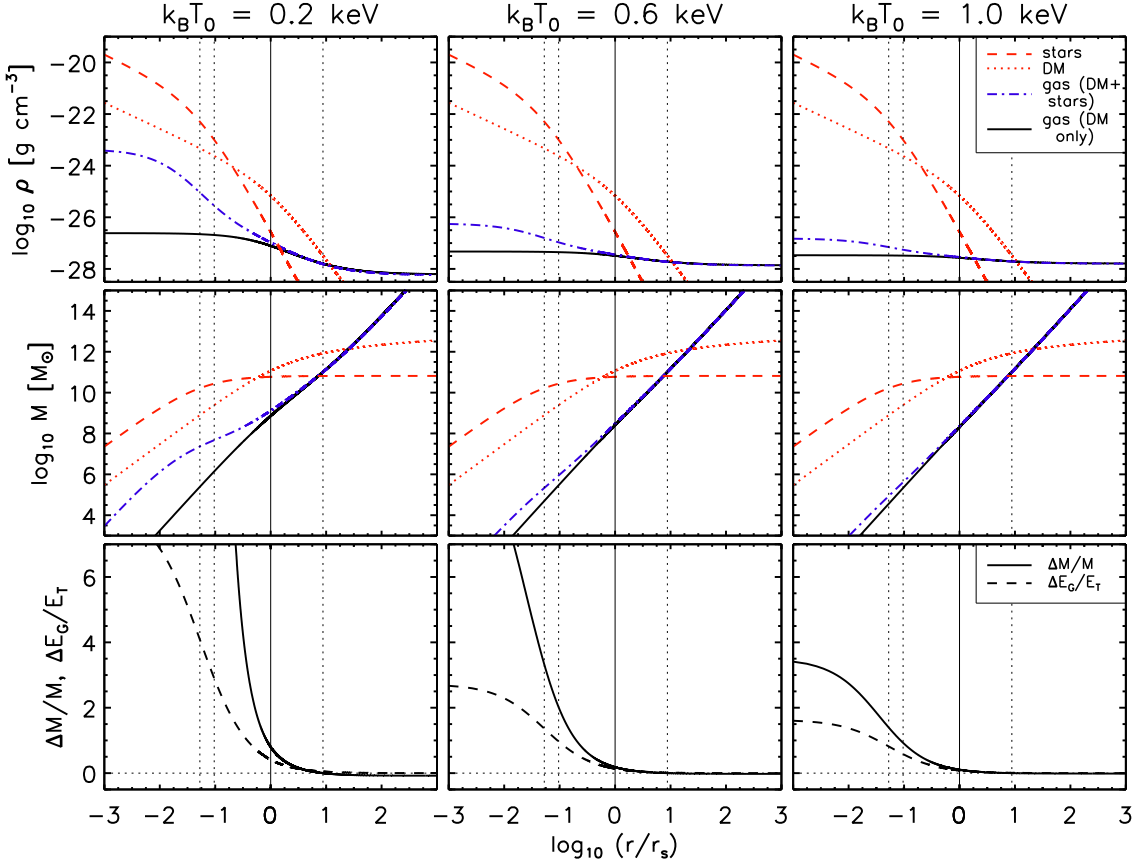}
\end{center}
\caption{Mass density, enclosed mass, and fractional gas mass change as a function of radius, for an elliptical galaxy with DM (NFW profile), stars (Hernquist profile), and isothermal gas. For the fiducial model (central panels), whose parameters are listed in Table \ref{tab:Parameters}, and for two models with lower (left panels) and higher (right panels) gas temperature, we plot in the top (middle) panels, the mass density (enclosed mass) profiles of DM (red, dotted curve), stars (red, dashed curve), and isothermal gas (blue, dot-dashed curve) in HE with DM and stars. The black, solid curve represents the isothermal gas in HE with DM only. In the bottom panels, we plot (solid curve) the fractional change of enclosed gas mass between the case when stars are included in the total gravitational potential, and the case when they are not; we also plot (dashed curve) the ratio between the change in gravitational energy of a gas particle, due to the addition of the stellar component, and its thermal energy. The four vertical lines denote the stellar scale radius $r_*$, the effective radius $r_e$, the NFW scale radius $r_s$ (solid line), and the virial radius $r_{vir}$, from left to right, respectively. The inclusion of the stellar component dramatically alters the gas profile at the very centre of the early-type galaxy, with the gas being more centrally concentrated, due to a deeper and steeper gravitational potential well. The effect of gas concentration is more pronounced for lower values of the gas temperature.}
\label{fig:Dependence_on_T}
\end{figure}

\begin{figure}
\begin{center}
\includegraphics[width=11.5cm,angle=90]{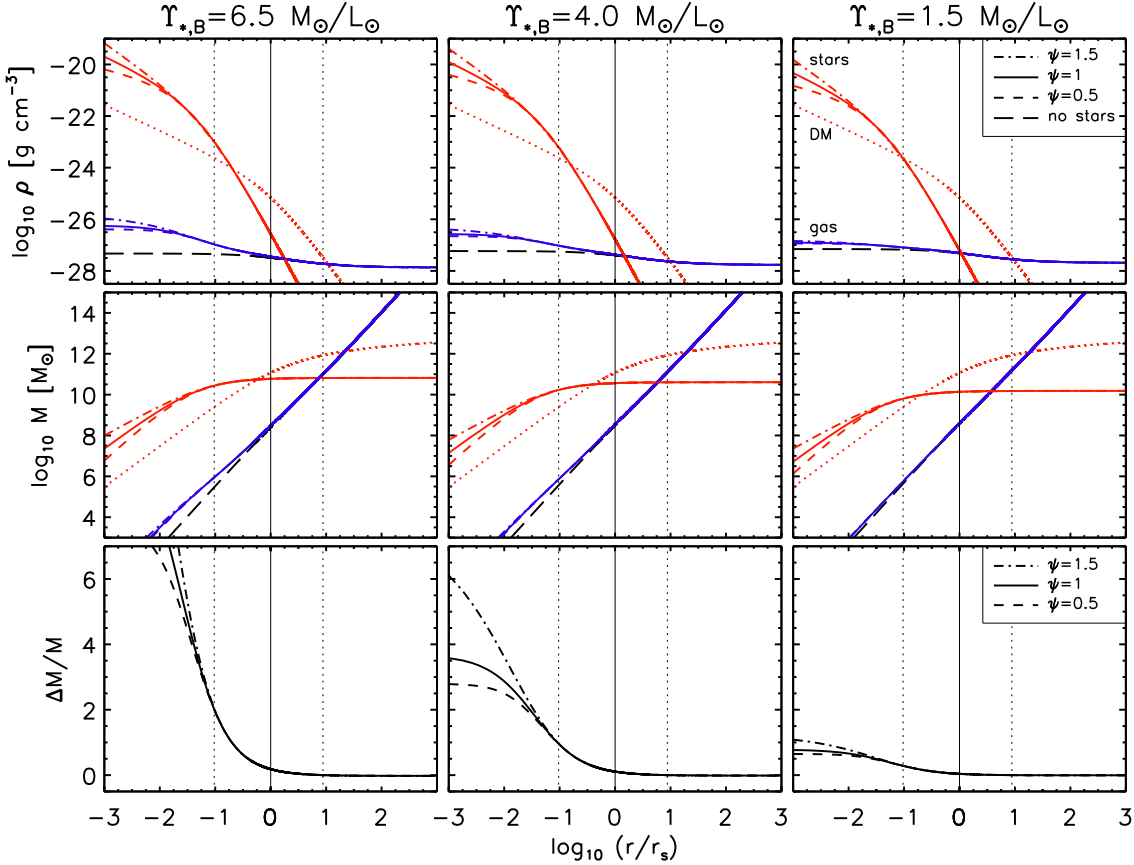}
\end{center}
\caption{Mass density, enclosed mass, and fractional gas mass change as a function of radius, for an elliptical galaxy with DM (NFW profile), stars (Dehnen profiles), and isothermal gas. For the fiducial model (with $\Upsilon_{*,B}=6.5$ M$_{\odot}/$L$_{\odot}$, $f_{star}/f_b \simeq 0.39$; left panels), whose parameters are listed in Table \ref{tab:Parameters}, and for two models with lower stellar baryon fraction ($\Upsilon_{*,B}=4.0$ M$_{\odot}/$L$_{\odot}$, $f_{star}/f_b \simeq 0.24$, central panels; and $\Upsilon_{*,B}=1.5$ M$_{\odot}/$L$_{\odot}$, $f_{star}/f_b \simeq 0.09$, right panels), we plot in the top (middle) panels, the mass density (enclosed mass) profiles of DM (red, dotted curve), stars, and isothermal gas in HE with DM and stars. The stellar (gas) profiles are the three upper, red (lower, blue) bundled curves, for three values of the Dehnen parameter: $\psi=0.5$ (dashed curve), $\psi=1$ (solid curve), and $\psi=1.5$ (dot-dashed curve). The black, long-dashed curve represents the gas profile when the stellar component of the total gravitational potential is neglected. In the bottom panels, we plot the fractional change of enclosed gas mass between the case when stars are included in the total gravitational potential, and the case when they are not, for the same three values of the Dehnen parameter. The three vertical lines in each panel denote the effective radius $r_e$, the NFW scale radius $r_s$ (solid line), and the virial radius $r_{vir}$, from left to right, respectively. The effect of gas concentration is more pronounced for higher values of the stellar mass fraction and for higher values of the Dehnen parameter.}
\label{fig:Dependence_on_stars}
\end{figure}

\begin{figure}
\begin{center}
\includegraphics[width=10.9cm,angle=90]{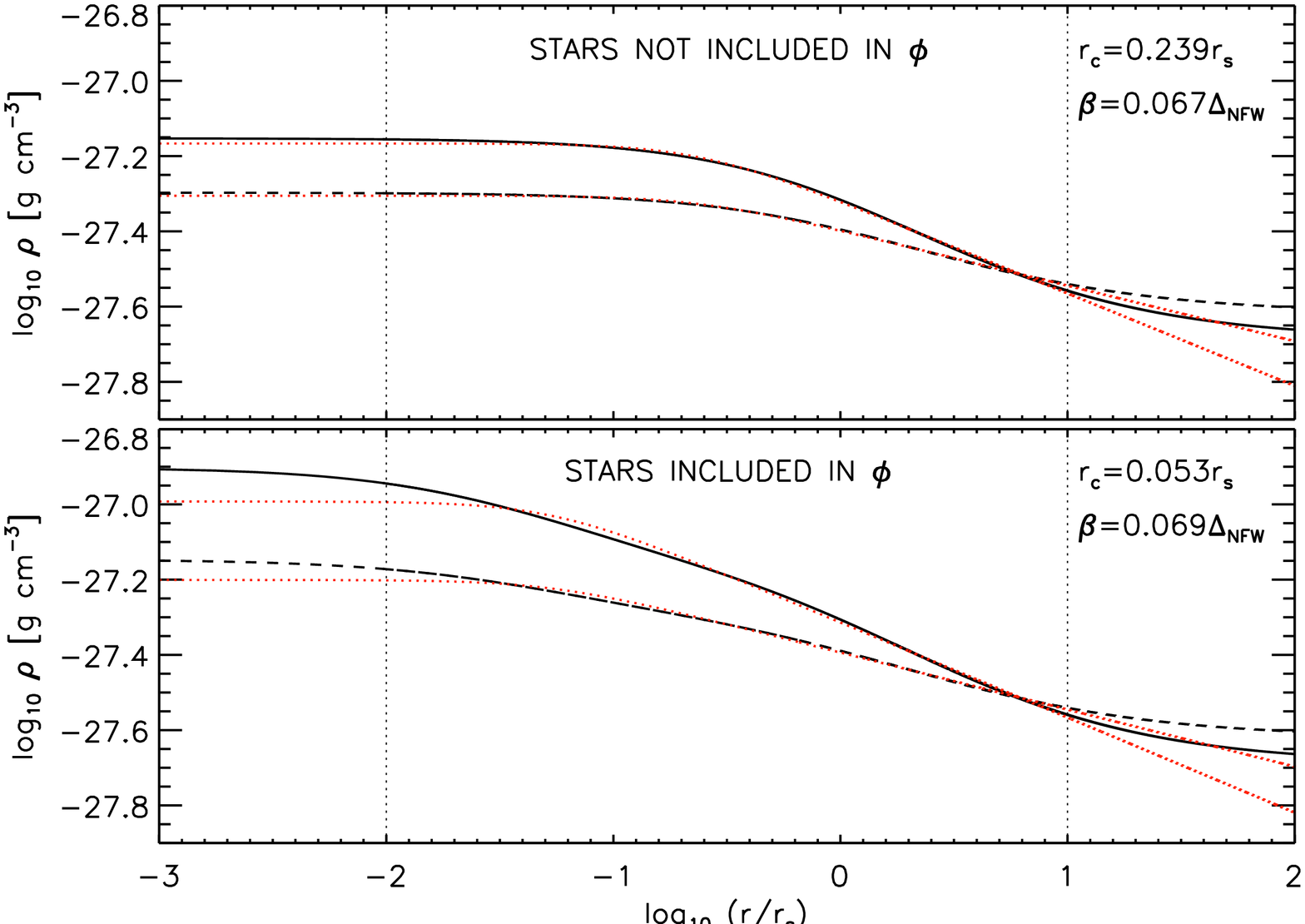}
\end{center}
\caption{For an elliptical galaxy with DM (NFW profile), stars (Hernquist profile), and isothermal gas, we plot the equilibrium gas density profiles excluding (top panel) and including (bottom panel) the effects of stars on the total gravitational potential, for a B-band stellar mass-to-light ratio $\Upsilon_{*,B}=1.5$ M$_{\odot}/$L$_{\odot}$ ($f_{star}/f_b \simeq 0.09$), for two different gas temperatures: $k_BT_0=0.6$ keV (black, solid lines) and $k_BT_0=1$ keV (black, dashed lines). The dotted, red curves are the best-fitting $\beta$-models in the $0.01\,r_s<r<10\,r_s$ range, denoted by the two dotted vertical lines. The inclusion of stars significantly affects the relation between the $\beta$-model scale radius and the NFW scale radius. For higher stellar mass fractions (not shown here), a $\beta$-model is unable to accurately reproduce our solution. When the observed surface brightness profile of an isolated elliptical galaxy is found to be well fitted by a $\beta$-model, the stellar mass fraction cannot be much greater than $\sim$10\%.}
\label{fig:Beta_model_comparison}
\end{figure}

In Fig. \ref{fig:Dependence_on_T} we plot the density (top panels) and enclosed mass (middle panels) profiles of DM, stars, and isothermal gas, for the fiducial model (with $k_BT_0=0.6$ keV; central panels) and for the lower and higher temperature models (with $k_BT_0=0.2$ keV, left panels; and $k_BT_0=1$ keV, right panels), with and without the stellar component in the gravitational potential. We focused on the isothermal solution because the radial range for which the polytropic solution is non-negative is limited\footnote{With the chosen parameters, the threshold temperature value above which the polytropic solution is physically meaningful at all radii is $k_BT_0 \sim 1$ keV.}. The inclusion of the stellar component dramatically alters the equilibrium profiles for the gas at the very centre of an early-type galaxy, with the gas being more centrally concentrated, due to a deeper and steeper gravitational potential well. The presence of the stellar component is thus explicitly taken into account and is manifested in the modification to the gas profile in the central region, in contrast to the case when stars are not included in the total gravitational potential. In the bottom panels of Fig. \ref{fig:Dependence_on_T}, we plot the fractional change of enclosed gas mass $\Delta M/M$ as a function of radius: for the fiducial model, in the central panel, the gas concentration ratio between the \textit{new} (i.e. including stars in the gravitational potential) and the \textit{old} (i.e. excluding stars) enclosed gas mass is $[1+\Delta M/M] \equiv f_{conc} \sim 4$ at the stellar scale radius $r_*$, $\sim 3$ at the effective radius $r_e$, and $\sim 1.2$ at the NFW scale radius\footnote{The fit for the effective radius can be considered uncertain to a factor of 2 (Mamon \& {\L}okas 2005a): for this reason, we multiplied and divided $r_e$ by 2, fixing all other parameters, and found that the effect of gas concentration increases as the effective radius decreases, as expected.}, $r_s$. We note that the effect of gas concentration increases with decreasing temperature, because the ratio between the change in gravitational energy of a gas particle, due to the addition of the stellar component, and its thermal energy, decreases with increasing temperature, as is shown in the bottom panels of Fig. \ref{fig:Dependence_on_T}. We also point out that our gas (density and enclosed mass) profile is very different from that found in Mamon \& {\L}okas (2005b). They assume a fixed $\beta$-model to describe the isothermal gas in the galaxy, and do not solve the HE equation directly. Note that, by construction, the enclosed gas mass at the virial radius is the same.

It is worth pointing out that the value of $f_{star}$ of the fiducial model corresponds to a value of stellar baryon fraction $f_{star}/f_b \simeq 0.39$, which is higher than that inferred from recent gravitational lensing studies, for which $f_{star}/f_b \simeq 0.1$ (e.g. Hoekstra et al. 2005; Mandelbaum et al. 2006; Heymans et al. 2006; Lagattuta et al. 2009; see also Fukugita, Hogan \& Peebles 1998). The discrepancy is likely due to different values of mass-to-light ratios and of mass-to-light ratio bias. Lagattuta et al. (2009), for example, use a V-band virial mass-to-light ratio $\Upsilon_V=210$ M$_{\odot}/$L$_{\odot}$ and a V-band stellar mass-to-light ratio $\Upsilon_{*,V}=2.8$ M$_{\odot}/$L$_{\odot}$. Using these values, we would obtain $f_{star}/f_b \simeq 0.1$. For this reason, we compare the gas radial profiles for three different values of $\Upsilon_{*,B}=1.5, 4.0, 6.5$ M$_{\odot}/$L$_{\odot}$, which correspond to stellar mass fractions $f_{star}=0.015, 0.04, 0.065$, and to stellar baryon fractions $f_{star}/f_b \simeq 0.09, 0.24, 0.39$, respectively. For these values of $\Upsilon_{*,B}$, we also vary the Dehnen parameter to explore dependencies on stellar profiles, choosing values of $\psi=0.5$, 1 (Hernquist 1990), and 1.5 (the ``best'' value in Dehnen 1993).

Fig. \ref{fig:Dependence_on_stars} shows the density (top panels) and enclosed mass (middle panels) profiles for the three chosen values of $\Upsilon_{*,B}$ and of $\psi$, in the same way we plotted Fig. \ref{fig:Dependence_on_T}, for a system with $k_BT_0 = 0.6$ keV. The effect of gas concentration is obviously more pronounced for higher values of the stellar mass fraction and for higher values of the Dehnen parameter, as it is shown in the bottom panels. However, the variation of the Dehnen parameter does not cause significant changes in the enclosed gas mass, except in the very inner-most regions of the galaxy ($r < r_e/10$): for example, for the $\Upsilon_{*,B}=6.5$ M$_{\odot}/$L$_{\odot}$ case, the enclosed gas mass at the effective radius varies by less than 1\% when changing the Dehnen parameter from $\psi=1$ to $\psi=0.5$, or to $\psi=1.5$. On the other hand, the change in stellar fraction is much more evident: for example, for the Hernquist case ($\psi=1$), the ratio between the enclosed gas masses at the effective radius $r_e$ varies from $\sim$3 in the $\Upsilon_{*,B}=6.5$ M$_{\odot}/$L$_{\odot}$ case, to $\sim$2 in the $\Upsilon_{*,B}=4.0$ M$_{\odot}/$L$_{\odot}$ case, to $\sim$1.3 in the $\Upsilon_{*,B}=1.5$ M$_{\odot}/$L$_{\odot}$ case.

\subsection{Comparison with the $\beta$-model}

By comparing the theoretically derived equilibrium solution for an isothermal gas to the observationally inspired $\beta$-model profile (Cavaliere \& Fusco-Femiano 1976, 1978), it is possible to derive relations between observed properties of the hot gas and the parameters of the underlying DM gravitational potential of a given system. Makino et al. pursue this (1998; see also Suto et al. 1998 and Wu \& Xue 2000 for other DM profiles) for the case of clusters of galaxies with a DM halo described by an NFW profile. They fit  their HE analytic solution for an isothermal gas (our equation \ref{eq:Makino_sol}) to a generic $\beta$-model profile given by

\begin{equation}
\rho_{\beta}(r) = \frac{\rho_{0\beta}}{[1+(r/r_c)^2]^{3\beta/2}} = \frac{\rho_0(A_0+A_1\Delta_{NFW})}{[1+[r/(A_2r_s)]^2]^{3A_3\Delta_{NFW}/2}},
\end{equation}

\noindent where $\rho_{0\beta}$, $r_c$, and $\beta$ are the central density, the scale radius, and the characteristic slope parameter of the $\beta$-model, respectively, and $A_i$ (for $i=$ 0-3) are fitting parameters, in the $0.01\,r_s<r<10\,r_s$ range, obtaining $r_c \simeq 0.22 \,r_s$ and $\beta \simeq 0.067 \,\Delta_{NFW}$, for a wide range of values of $\Delta_{NFW}$. We performed a non-linear least squares fitting with the IDL routine MPFIT (Markwardt 2009), using our own values of $\Delta_{NFW}$, and obtained similar results in the case when stars are not included in the total gravitational potential, regardless of the value of the stellar mass fraction and of the gas temperature: $r_c \simeq 0.239 \,r_s$ and $\beta \simeq 0.067\,\Delta_{NFW}$ in the $0.01\,r_s<r<10\,r_s$ range. Note that, even though the $\beta$-$\Delta_{NFW}$ relation is independent of temperature, the value of the characteristic slope parameter $\beta$ does depend on temperature through $\Delta_{NFW}$.

Now we go a step further and compare the $\beta$-model profile to the isothermal HE gas profile in the case when stars are included in the gravitational potential, given by equation \eqref{eq:Capelo_iso}. Since we found that there is no strong dependence on the Dehnen parameter, we focus on the Hernquist case ($\psi = 1$) for simplicity. Also, we choose the stellar B-band mass-to-light ratio to be $\Upsilon_{*,B}=1.5$ M$_{\odot}/$L$_{\odot}$, as the inferred stellar baryon fraction $f_{star}/f_b \simeq 0.09$ is closer to what found from gravitational lensing observations (e.g. Hoekstra et al. 2005), and because higher stellar mass fraction systems do not fit the $\beta$-model profiles as well. The fact that it is indeed possible to fit X-ray observations of early-type galaxies with $\beta$-model profiles (e.g. O'Sullivan, Ponman \& Collins 2003), is an additional indication that the stellar baryon fraction of such galaxies cannot be much higher than $\sim$10\%.

In Fig. \ref{fig:Beta_model_comparison} we show the equilibrium gas density profiles for an isothermal gas in a system with NFW DM and Hernquist stars, for two different temperatures, including and excluding the effects of stars on the total gravitational potential. All HE gas profiles are fitted to $\beta$-model profiles in the $0.01\,r_s<r<10\,r_s$ range. We chose this particular range to be able to make consistent comparisons with the results of Makino et al. (1998), and also because it is typically difficult to obtain precise X-ray measurements at larger radii (due to the low surface brightness of the gas) and at smaller radii (due to resolution limitations). We note that, whereas the relation between $\beta$ and $\Delta_{NFW}$ does not change significantly ($\beta \simeq 0.067\,\Delta_{NFW}$ excluding stars, and $\beta \simeq 0.069\,\Delta_{NFW}$ including stars), the $r_c$-$r_s$ relation between the two scale radii varies substantially, from $r_c \simeq 0.239 \,r_s$, when stars are not included in the potential, to $r_c \simeq 0.053 \,r_s$, when stars are included. We note that the particular constant of proportionality between $r_c$ and $r_s$ depends on the assumed stellar mass fraction. Using the $r_c$-$r_s$ relation can therefore lead to an underestimate of the NFW scale radius, when stars are not taken into account. Defining $f_{r_s} \equiv r_{s-new}/r_{s-old}$ as the ratio between the $\beta$-model inferred value of $r_s$ when stars are included in the gravitational potential (the \textit{new} value), and when they are not (the \textit{old} value), we have $f_{r_s} \sim 5$. Assuming the virial mass $M_{vir}$ and radius $r_{vir}$ of the system are known, we then have the ratio of concentration parameters $c_{vir-new}/c_{vir-old} = 1/f_{r_s} \sim 0.2$. Therefore, the enclosed DM mass at $r_{s-new}$ would be overestimated by a factor of $[\ln(1+f_{r_s}c_{vir-new})-f_{r_s}c_{vir-new}/(1+f_{r_s}c_{vir-new})]/[\ln(1+c_{vir-new})-c_{vir-new}/(1+c_{vir-new})] \sim 2$.

\section{Conclusions}\label{sec:Conclusions}

We solve the HE equation analytically for a system with DM described by an NFW profile, and stars described by a Dehnen profile, for the two common barotropic cases (the isothermal case and the polytropic case). We thus extend previous calculations which only considered the potential due to DM. In doing so, we do assume spherical symmetry, and we neglect the effects of gas self-gravity and adiabatic contraction. We compare the newly obtained equilibrium solutions as a function of stellar mass fraction, central gas temperature, and the Dehnen parameter, to the case when the stellar component is included in the total mass budget but not in the total gravitational potential. We determine that neglecting the effects of stars leads to an underestimate of the enclosed gas mass in the inner regions of the galaxy. We then present new relations between $\beta$-model and DM parameters, and find that the inclusion of the stellar component is necessary to avoid an underestimate of the NFW scale radius. In particular:

\begin{itemize}
\item For a system with NFW DM and Hernquist stars at redshift $z=0.5$, with virial mass $M_{vir}=10^{12}$ M$_{\odot}$, and central temperature $k_BT_0=0.6$ keV, we find in the isothermal case that the gas concentration ratio $f_{conc}$ between the \textit{new} (i.e. including stars in the gravitational potential) and the \textit{old} (i.e. excluding stars) enclosed gas mass, can be as high as $\sim$3 at the effective radius, and as high as $\sim$1.2 at the NFW scale radius, for a stellar baryon fraction $f_{star}/f_b \simeq 0.39$.
\item The effect of gas concentration decreases with decreasing stellar baryon fraction as expected, with $f_{conc}(r_e)$ varying from $\sim$3 in the $f_{star}/f_b \simeq 0.39$ case, to $\sim$2 in the $f_{star}/f_b \simeq 0.24$ case, to $\sim$1.3 in the $f_{star}/f_b \simeq 0.09$ case.
\item On the other hand, the effect of gas concentration is virtually independent of the Dehnen parameter, except in the very inner-most regions of the galaxy ($r < r_e/10$), with the value of $f_{conc}(r_e)$ changing by $\sim$1\% when varying the Dehnen parameter in the $0.5 < \psi <1.5$ range.
\item We further determine that the effect of gas concentration decreases with increasing temperature, as expected.
\item We find similar results for the polytropic case at higher temperatures, where the solution is physically meaningful. In fact, for a fixed central gas temperature, the effect of gas concentration is more pronounced than in the isothermal case, because the polytropic gas has a lower mean temperature.
\item We derive new relations between observed properties of the hot gas, described via a $\beta$-model, and the parameters of the underlying DM gravitational potential of early-type galaxies, which avoid an underestimate of the NFW scale radius by $\sim$80\%, and an overestimate of the enclosed DM mass at $r_s$ by a factor of $\sim$2, assuming a 10\% stellar baryon fraction.
\item Models with higher stellar mass fraction do not fit $\beta$-models well, further indicating that the stellar baryon fraction of isolated elliptical galaxies whose observed surface brightness profile is well fitted by a $\beta$-model cannot be much higher than $\sim$10\%.
\end{itemize}

These new analytic solutions will be useful for the interpretation of X-ray data of elliptical galaxies and in particular, will help derive unbiased estimates of the baryon mass fraction as a function of radius. Moreover, the formalism in equations (\ref{eq:Capelo_iso}-\ref{eq:Capelo_poly}) provides a new prescription for setting up more realistic initial conditions for simulations of galaxy formation (e.g. Di Matteo et al. 2005; Springel et al. 2005; Hopkins et al. 2006).

\section*{Acknowledgments}

PRC is grateful for many helpful discussions with Andrew Szymkowiak.

\end{document}